# Refining Ray-Tracing Accuracy and Efficiency in the Context of FRMCS Urban Railway Channel Predictions

Romain Charbonnier, Thierry Tenoux, Yoann Corre
*SIRADEL*
Saint-Grégoire, France
rcharbonnier@siradel.com

*Abstract*— The upcoming roll-out of the new wireless communication standard for wireless railway services, FRMCS, requires a thorough understanding of the system performance in real-world conditions, since this will strongly influence the deployment costs and the effectiveness of an infrastructure planned for decades. The virtual testing of the equipment and network performance in realistic simulated scenarios is key; its accuracy depends on the reliability of the predicted radio channel properties. In this article, the authors explain how they are evolving a ray-tracing (RT) tool to apply it to the specific case of simulating the radio link between the FRMCS fixed infrastructure and an antenna placed on the roof of a train moving in an urban environment. First, a dynamic version of the RT tool is used to capture the rapid variations of all channel metrics; a compromise is sought between computation time and accuracy. Besides, a hybridization of RT and physical optics (PO) allows the integration of objects near the track, such as catenary pylons, into the simulation. A case study shows that the scattering by metallic pylons brings a significant contribution.

*Keywords—ray-tracing, railway, physical optics, FRMCS*

## I. Introduction

The new communication standard FRMCS (Future Railway Mobile Communication System) based on 5G NR technology is intended for private railway communications. It is set to replace the currently operational GSM-R standard, enabling higher throughput communications such as video and low latency. While the specification work is still ongoing, several countries have already planned experiments and are expected to begin deploying the new infrastructure within a few years. Two frequency bands have been allocated by CEPT (European Conference of Postal and Telecommunications Administrations): 2×5 MHz in the 900 MHz spectrum; and 10 MHz at 1900 MHz. A crucial challenge for future FRMCS operators is the proper network sizing and the optimal placement of base stations (possibly several thousand for a country) to minimize costs while ensuring continuous coverage and high reliability. The simulation of the radio channel, which is the subject of this article, is a key element for this optimization work. Indeed, frequency-selective MIMO channel predictions may contribute in the Research and Development phase to adjusting the 5G NR physical layer algorithms (e.g., schedulers) to the specificities of rail communications, or defining the most efficient antenna networks, etc. And when performed deterministically, i.e. based on real cartographic data, the radio channel prediction allows testing and choosing the best location for base stations.

The simulation of the FRMCS radio channel presents several challenges. Firstly, in an obvious way, the channel properties are highly dynamic due to the speed of the train. For example, if we want to simulate the performance of an FRMCS system over a period of 30 seconds for a train traveling at 300 km/h, the channel variations must be predicted over a distance of 2.5 km. Additionally, between two timestamps separated by only 1 ms (a typical timescale for the scheduler), the train travels 8.3 cm (more than λ/2 at 1900 MHz), the fades experienced by received signal are more or less correlated depending on the local coherence distance. A tool is needed that can accurately and efficiently predict the continuous combination of small and large-scale channel variations. Secondly, the encountered environments are very particular, such as trenches in rural areas, tunnels, passages through stations, and gaps in dense urban environments; transitions can be complex. It is necessary to have either channel models adapted to each scenario or a versatile one. Thirdly, the nature of interactions generating significant multipath contributions is very varied, including (from largest to smallest) mountains, buildings, bridges, or the many objects distributed along the tracks. Finally, to our knowledge, to date, there is no model that has been validated by comparison with channel measurements in the FRMCS 1900 MHz band. Note this latter challenge is critical but will not be discussed further in this article.

As reported in [1], several statistical models based on the simple TDL (Tapped Delay Line) structure have been derived from broadband measurements in the high-speed rail (HSR) context. A Doppler spectrum may sometimes be associated. But angular information is missing from most of the proposed empirical models, except if considering generic rural CDL (Cluster Delay Line) formulation by 3GPP or ITU. There is obviously a lack of statistical models addressing specifically the MIMO railway channel scenarios. Besides, the ray-tracing (RT) approach has the advantage it provides all the channel properties required for time-variant frequency-selective MIMO predictions, in a site-specific manner. Several works over the last ten years have demonstrated the interest and applicability of RT for railway scenarios, either at sub-6GHz or in the millimeter-wave band. RT is confirmed in [2] to give fast deterministic multi-path-predictions in diverse scenarios, including the urban, cutting and viaduct situations. Ambition of the authors of the present article is to predict accurate train-to-infrastructure time-variant spatially-consistent MIMO channel traces, with fine time resolution (possibly 1 ms), based on the Volcano ray-tracing tool [3]. This model did already support a wide range of outdoor and indoor radio scenarios, which answers part of the requirements. But improvements were necessary to achieve time-efficient dynamic simulations and integrate new impacting scatterers.

The optimization of RT-based time-varying simulations is currently attracting a lot of interest, particularly for vehicular use cases. For example, the Dynamic-RT approaches presented in [4] and [5] extrapolate the local variations in channel properties from a discrete RT prediction, using analytical formulas to track the evolution of the interaction points. They are applicable to both antennas and moving objects. However, perfect continuity between two successive calculation intervals does not seem necessarily guaranteed. We have implemented an approach that is slightly different; the propagation paths are tracked from one discrete RT prediction to another, then their geometric properties and losses are interpolated.

Additionally, we want to integrate objects placed near the train, which are likely to create strong interactions but are not compatible with the calculation methods usually associated with ray-tracing, i.e., Geometrical Optics (GO) and Uniform Theory of Diffraction (UTD), due to their size and/or shape. We choose to hybridize the ray-tracing method with a scattering calculation using Physical Optics (PO) which allows for greater flexibility. An originality of our approach lies in the possibility of combining discrete scattering with other existing RT interactions along the same propagation path. The technique is applied to catenary pylons that are systematically found along the tracks.

The article is organized as follows. Section II explains the principles behind the evolution of the RT tool. Section III details the settings of the considered FRMCS dynamic scenario, in a dense urban area at frequency 1900 MHz. Section IV analyzes the simulation results. Conclusions are drawn in Section V, along with some perspectives for future work.

## II. SIMULATION PRINCIPLES

### A. Ray-tracing model

The RT tool used in this work is VolcanoUrban. Over the past twenty years, it has been calibrated and validated by comparison with numerous field measurements at the FR1 and FR2 frequencies used by cellular communication networks. It is based on the launching technique described in [5], which allows for the rapid construction of 3D paths specially when the number of reception points is large, for example, for points distributed on a grid or along a road. The propagation paths result from the combination of several interactions with buildings: reflection on facades; diffraction on vertical edges; and/or diffraction by roofs. The inaccuracies induced by the ray-launching are corrected afterward to place the interaction points at the exact position determined by the image theory. For FRMCS simulation at link or system level, we aim to predict the variations of the multi-dimensional radio channel with strict continuity, temporal resolutions on the order of milliseconds, over a period of several seconds or even tens of seconds. If the environment is static and only the train antenna is moving, then the ray-launching can consider a reception point at each timestamp, thus handling this scenario with reasonable computational effort. However, we aim to accelerate further the calculations and approach real-time performance.

### B. Ray-tracing interpolation

In the problem addressed in this article, RT is applied to a scenario where a mobile antenna travels along a linear trajectory through a static environment at constant or varying speeds. The use of RT deterministic time-correlated multi-paths allows for obtaining time-varying wideband MIMO coefficients that are consistent across all spatial, temporal, and frequency dimensions. However, when targeting time resolutions as low as 10 ms or 1 ms, the computational effort for RT becomes significant. The authors propose using ray-path interpolation.

Exact ray-paths are computed using the conventional RT approach at only a few regularly spaced positions along the antenna trajectory, such as points Pi, Pi+1, and Pi+2 shown in Fig. 1. The separation between exact RT calculations must be less or similar to the channel stationarity distance. Each computed ray is described by its multi-polar complex field, delay, departure/arrival angles, and complete 3D geometry. Rays are tracked from one position to another. For example, the rays represented by the green arrow at positions Pi, Pi+1, and Pi+2 in Fig. 1 are similar in that they undergo the same sequence of physical interactions with the same obstacles. This ray-path is assumed to exist at any timestamp where the mobile antenna is located between Pi and Pi+2; its received field amplitude, 3D trajectory, angles, and delay are obtained from simple interpolation laws. Variations in the received field phase are deduced from the delay. A random birth or death process is executed when a new ray-path must appear (e.g., the blue path in Fig. 1) or disappear (e.g., the orange path). This technique ensures a smooth and realistic evolution of the channel properties, regardless of the required time resolution or the simulated travel duration. Of course, more abrupt variations occur at timestamps where a dominant ray is born or dies.

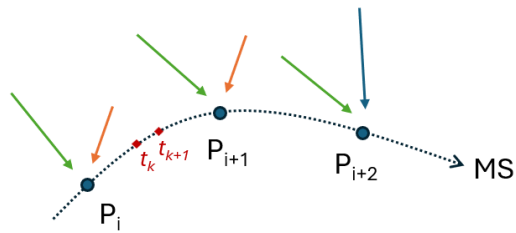

Fig. 1. Ray paths (colored arrows) predicted at regular steps along the mobile station (MS) trajectory

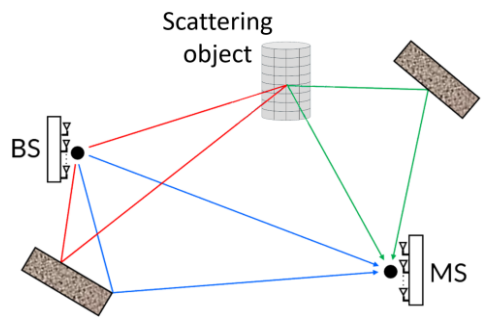

Fig. 2. Kinds of predicted ray paths: specular (blue); incident (red); or scattered (red).

In the following of this article, we evaluate the errors due to interpolation for an urban FRMCS scenario, and look for RT resolutions (the distance between two exact calculations) that offer interesting trade-offs between computation time and accuracy. The errors are estimated in all dimensions of the radio channel, by comparing to a systematic RT calculation.

### C. Hybridization with Physical optics

The scattering on small objects placed along the track (those higher than the train) certainly plays a significant role in the FRMCS channel characteristics in open environments, as these interactions are sometimes the only ones providing diversity in delay and angle. In this article, we use simulation to observe their contribution in an urban environment and assess their importance relative to other multipaths. To achieve this, the conventional RT approach is complemented with the prediction of discrete scatterers. PO is used to calculate the scattered field instead of GO or UTD, since it is more accurate when the surfaces of the interaction object are non-flat or small relative to their distance from the antennas.

TABLE I. SIMULATED RAILWAY SCENARIO

| Simulation parameters | Value |
|---|---|
| Central frequency | 1900 MHz |
| BS height | 20.5 m |
| BS antenna | Dual-polar V/H omni-directional |
| BS antenna gain | 0 dBi |
| BS transmit power | 43 dBm |
| Distance between BS and track | 20.0 m |
| Train speed | 100 km/h |
| UE height | 4.5 m |
| UE antenna | Dual-polar V/H omni-directional |
| UE travelling duration | 1 min / 1.67 km |

TABLE II. RAY-TRACING SETTINGS

| Ray-tracing (RT) parameters | Value |
|---|---|
| Nb lateral reflections | 2 |
| Nb lateral diffractions | 1 |
| Nb rooftop diffractions | No restriction |
| Channel update | Every 10 ms / 27.8 cm |
| Interpolation / RT resolution | From 50 ms to 500 ms / From 1.4 m to 13.9 m |

TABLE III. SCATTERING PREDICTION SETTINGS

| Scattering by pôles | Value |
|---|---|
| Height of poles | 8.2 m |
| Radius of poles | 0.375 m |
| Shape of poles | Cylinder |
| Material | Metal |
| Meshing resolution | $\lambda/2$ |

The roles are distributed as follows: RT is used to calculate the channel contribution due to direct propagation and specular interactions, as illustrated by the blue paths in Fig. 2. RT also provides the trajectory of propagation paths incident on the scatterer (in red) and the trajectory of paths leaving the scatterer (in green). Then, the field of all possible scattered contributions is computed using PO by considering the given incident and scattered angles. The basic principles are similar to those previously presented by the authors for the simulation of Reconfigurable Intelligent Surfaces (RIS) [6]. The computation may involve discretizing the object into smaller facets and subsequently discretizing the PO integral [7] to achieve a reliable solution in both far- and near-field ranges.

In the presented study, a pylon is represented by a meshed, perfectly-conducting vertical cylinder. The mesh surfaces, i.e., small rectangular surfaces, are assumed to re-radiate the incident power following the bistatic analytical far-field Radar Cross-Section (RCS) given in [8]. The size of the facets is half a wavelength, thus making the PO approach valid in all considered situations.

III. SIMULATION SCENARIO

The radio channel is emulated for a typical urban HSR scenario at the FRMCS 1900MHz frequency, based on the actual Paris environment depicted in Fig. 3. The scenario features a macro base station mounted on a building rooftop, 20.5 meters above ground level and 20 meters from the railway track. The mobile antenna is positioned on the roof of a train, 4.5 meters above the ground, traveling at a constant speed of 100 km/h for a duration of 60 seconds. Note that 81% of the mobile antenna locations are in line-of-sight (LoS) or only obstructed by trees. All scenario parameters are detailed in Table 1.

The settings of the RT model are provided in Table 2. The performance of RT interpolation, in terms of computation time and accuracy, is evaluated by comparing it to a systematic RT computation performed every 10 ms (equivalent to 27.8 cm). Various RT resolutions, ranging from 50 ms (i.e., 1.4 meters) to 500 ms (i.e., 13.9 meters), are tested. At each 10 ms timestamp, the interpolated channel properties are compared to the (reference) systematic calculation results.

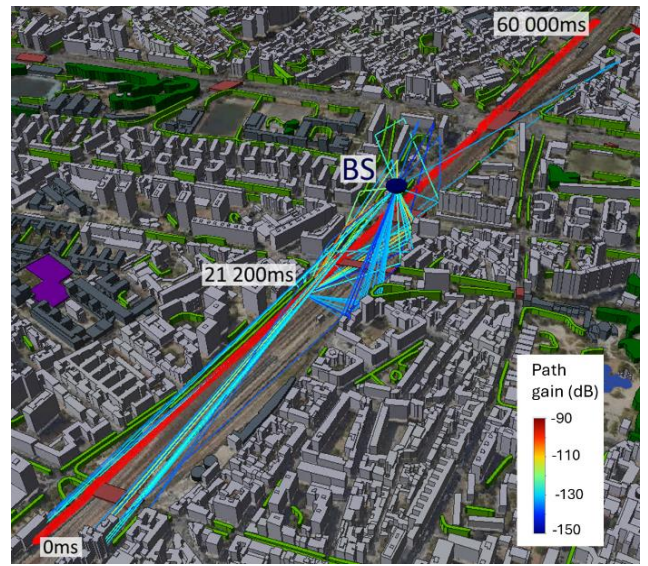

Fig. 3. Trajectory of the predicted mobile antenna (red line); location of the base-station (blue dot); and example of ray-paths at timestamp 21.2 s

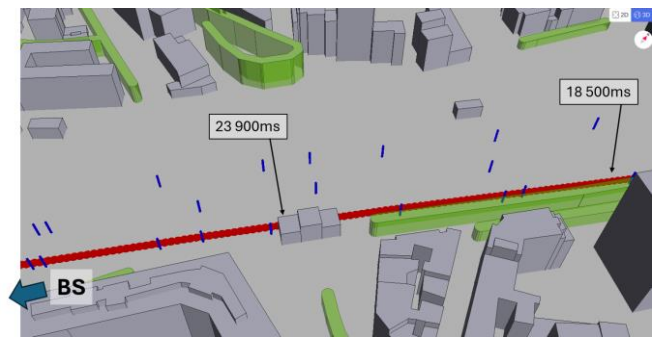

Fig. 4. Zoom on the area where the effect of pylons (blue cylinders) is evaluated.

Fig. 4 illustrates several catenary pylons added to a specific section of the scenario area, with their locations inspired by the real environment. Each pylon is assumed to be composed of a cylindrical metallic pole with a height of 8.2 meters. For our evaluation study, the contributions scattered by these poles are predicted for timestamps between 18.5 s and 23.9 s (covering a distance of 150 meters). More details on the scattering simulation settings are provided in Table 3.

## IV. RESULTS

### A. Adjustment of the RT interpolation interval

The degradation induced by the RT interpolation is characterized and tuned based on a multi-dimensional analysis. At each predicted timestamp, we calculate the error on the following metrics: received power VV; received power HV; received power HH; received power VH; mean delay; delay spread; mean horizontal angle of arrival (HAoA); horizontal angle of arrival spread; mean vertical angle of arrival (VAoA); vertical angle of arrival spread; mean Doppler shift; and Doppler shift spread. Fig. 5 shows the error statistical distribution of the VV and HV received power and the Doppler shift observed over the 6000 channel predictions. The received powers are predicted on a single carrier, resulting from the narrowband summation of multi-path complex fields, and thus undergo fast fading fluctuations. As expected, the errors in the VV power increase as the interpolation interval grows, but they mostly remain below 2 dB (in absolute value). The degradation in the HV power is significantly greater, with approximately 38% of timestamps experiencing at least a 2 dB error for an interpolation interval of 500 ms. This is due to the larger amplitude of cross-polar power fluctuations. Additionally, we observed that the error distributions for co-polar VV and HH are very similar, as are those for cross-polar HV and VH.

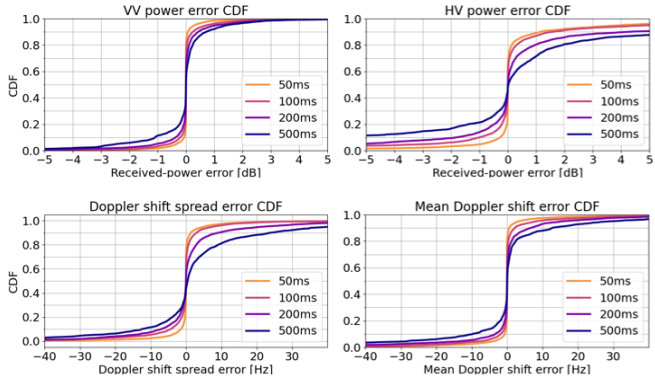

Fig. 5. Cumulative Distribution Function (CDF) of the interpolation errors

A normalized RMSE (Root Mean Square Error) is calculated for each considered metric:

$$NRMSE = \frac{RMSE}{Q90 - Q10} \quad (1)$$

where the RMSE is obtained from the 6000 predicted channels; Q10 and Q90 are respectively the 10% and 90% quantiles. The evolution of NRMSE's is plotted as a function of the interpolation interval in Fig. 6 and is analyzed together with the reduction in the normalized computation time showed in Fig. 7. The normalization factors for NMRSE are respectively 22.9 dB for VV power; 27.2 dB for HV power; 347 Hz for the mean Doppler shift; 102 Hz for the Doppler shift spread; 2.30 ms for the mean delay; 0.56 ms for the delay spread; 185° for the mean HAoA; 55° for the HAoA spread; and 8.7° for the mean VAoA. The VAoA spread is not reported in Fig. 6, since the normalization factor (Q90-Q10) is only 1.3°, thus even modest errors may lead to large NRMSE.

We observe that beyond interval 100 ms (equivalent to 2.8 meters), errors continue to increase significantly, while the gains in computation time become less and less interesting. At 100 ms, the NRMSE of a large majority of metrics is less than or equal to 10%, and the computation time is almost divided by 5. This interval offers an interesting compromise for a default value proposed by the tool. Then, depending on the aim and constraints of each use case, it can be adjusted with full knowledge of the impacts illustrated in Fig. 6 and 7.

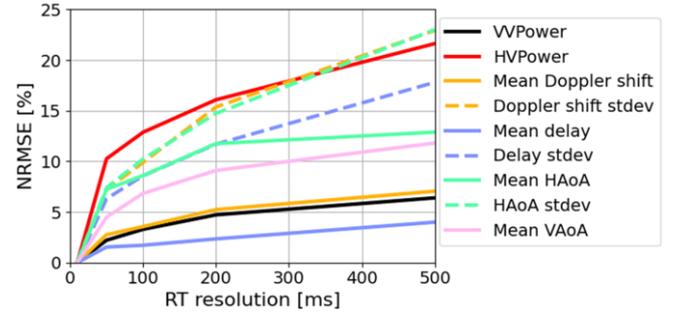

Fig. 6. Normalized prediction errors vs. Interpolation RT resolution

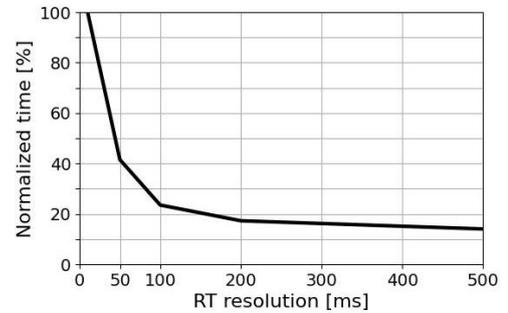

Fig. 7. Normalized computation time vs. Interpolation RT resolution

### B. Impact of catenary pylons on the radio channel

The simulation of the radio channel with catenary pylons is carried out over an interval of 4.4 seconds. The result is presented in the form of a time-varying channel impulse response (TV-CIR) for polarizations VV, with time resolution 100 ms. A raised cosine filter with a bandwidth of 100 MHz and a roll-off factor β=0.95 is assumed; the chosen bandwidth is ten times the one allowed for a FRMCS signal, to facilitate the visual discrimination between multi-paths. Fig. 8 displays the TV-CIR as a colored map with time on the x-axis and delay on the y-axis. In the top graph of Fig. 8, the trace of the direct path and paths traveling a distance very close to the direct one is clearly visible; this broad and dark red trace (high power) shows the lowest delay, which decreases almost linearly as the base station is approached by the mobile antenna. Other visible traces, with lower power levels, are due to multipaths with delays greater than 50 ns (equivalent to 15 meters). These paths result from interactions with buildings near the railway or scattering on a pylon, or a combination of both phenomena. The traces of indirect paths show an increase in delay as the train moves away from the last interaction point. The impact of contributions scattered by the pylons has been isolated in the bottom graph of Fig. 8; it obviously represents a significant part of the multipaths; and brings specific characteristics to the urban railway channel.

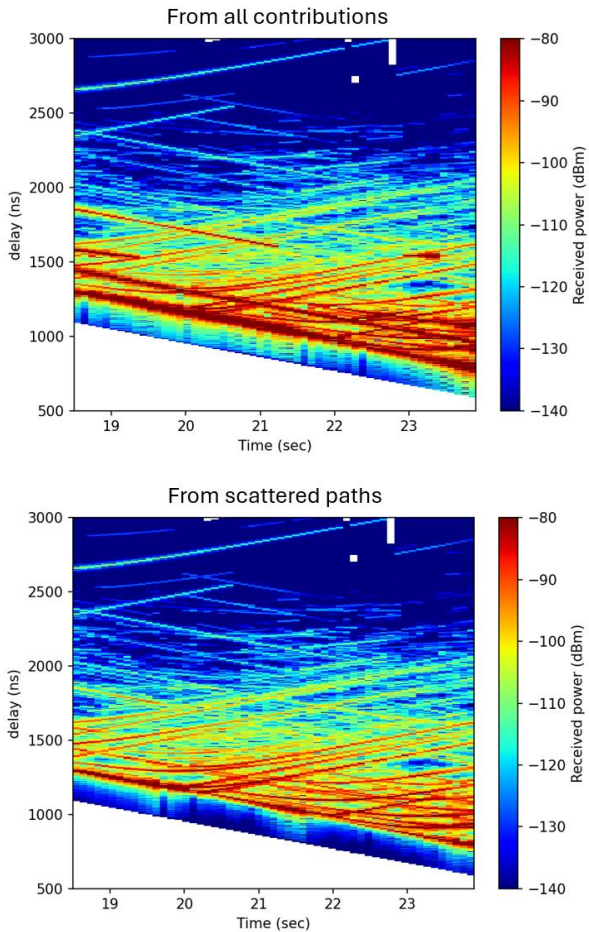

Fig. 8. Time-variant channel impulse response (CIR) from all contributions (top) or only the scattered paths (bottom)

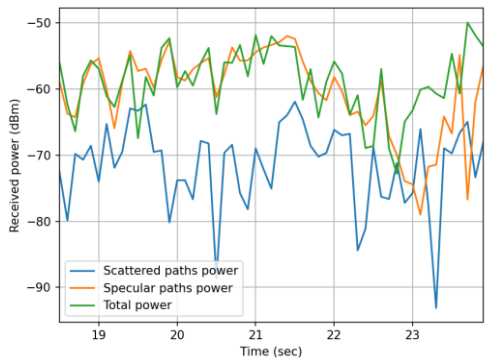

Fig. 9. Time evolution of the specular, scattered, and total power.

Fig. 9 shows the evolution of the received power on the carrier at the center of the signal band. This power fluctuates depending on the dominant paths, obstruction effects, and the phase of the multiple received fields. We compare i) the power of paths scattered by a pylon, ii) the power of the specular channel composed of paths interacting with buildings, and iii) the total power resulting from the sum of all contributions. The first is often 10-20 dB lower than that of the specular or total, except between 21.8 s and 23.5 s, where the direct path is obstructed by a building located nearby the track. Here, the presence of pylons in line-of-sight of both the base station and the mobile antenna allows the scattered contribution to be often less than 10 dB below the total power. In average over the entire simulated time interval, the specular power represents 70% of the total power; the scattered power represents 10%. The average delay spread measured from the direct and specular paths together is 69 ns; but increases to 77 ns when adding the pylons scattering effect.

## V. CONCLUSION

This article illustrates and evaluates two mechanisms to adapt a ray-tracing tool to FRMCS railway scenarios: the interpolation of multipaths to reduce computational effort; and the integration of discrete scatterers near the railway track through hybridization with physical optics. We integrated these techniques in the Volcano ray-tracing, and tested them on an urban scenario considering a fixed base station near the track and a mobile antenna installed on the roof of the train at 1900 MHz. We first observe that an interpolation interval of 100 ms is a good compromise for the dynamic simulation of specular contributions. Then, we note the importance of contributions scattered by catenary poles (when modelled as vertical metallic cylinders with diameter 70 cm) in the channel impulse response, and even in the NLoS total power. We believe that beyond this specific study, the same mechanisms and evaluation procedures presented in the article are relevant for a large number of railway and vehicular scenarios.

Analyses will continue by integrating the scattered contributions in the interpolation task and by testing different shapes and materials for the pylons. Furthermore, the new simulation tool will soon be compared to channel measurements collected by our partners as part of the projects mentioned in the Acknowledgment section; that will permit us to tune and validate the scattering model. The ultimate goal is to conduct reliable system performance studies for the evaluation of FRMCS technology.

ACKNOWLEDGMENT

This work is supported by the German-French innovation project 5G-RACOM under grant agreement 01MJ22015; and the French project 5G-REMORA under grant n° ANR-22-CE22-0015.

REFERENCES

[1] M. Berbineau, R. Behaegel, J.M. Garcia-Loygorr, R. Torrego, R. D'Errico, A. Sabra, Y. Yan, J. Soler, "Channel models for performance evaluation of wireless systems in Railway environments", *IEEE Access*, 2021.

[2] D. He et al., "Ray-tracing simulation and analysis of propagation for 3GPP high speed scenarios", 11th European Conference on Antennas and Propagation (EuCAP), Paris, March 2017.

[3] Y. Corre and Y. Lostanlen, "Three-Dimensional Urban EM Wave Propagation Model for Radio Network Planning and Optimization Over Large Areas," in *IEEE Transactions on Vehicular Technology*, vol. 58, no. 7, pp. 3112-3123, Sept. 2009.

[4] F. Quatresooz, S. Demey and C. Oestges, "Tracking of Interaction Points for Improved Dynamic Ray Tracing," in *IEEE Transactions on Vehicular Technology*, vol. 70, no. 7, pp. 6291-6301, July 2021.

[5] D. Bilibashi, E. M. Vitucci and V. Degli-Esposti, "Dynamic Ray Tracing: Introduction and Concept," 14th European Conference on Antennas and Propagation (EuCAP), Copenhagen, Denmark, 2020.

[6] G. S. Bhatia, Y. Corre, T. Tenoux and M. Di Renzo, "Exploring RIS Coverage Enhancement in Factories: From Ray-Based Modeling to Use-Case Analysis," 18th European Conference on Antennas and Propagation (EuCAP), Glasgow, UK, 2024.

[7] P. Pouliguen, L. Desclos, "A physical optics approach to near field RCS computations", *Annales des télécommunications*, vol. 51, no. 5-6, pp. 219-226, 1996.

[8] C. A. Balanis, *Advanced Engineering Electromagnetics*, Ed. Wiley, 2012.